\documentstyle[prl,aps]{revtex}
\begin{document}
\draft
\title{How much time does a tunneling particle spend
in the barrier region?}
\author{Aephraim M. Steinberg\footnote{Internet
aephraim@physics.berkeley.edu;
As of 1995, at NIST, Phys A167, Gaithersburg, MD 20899}}
\address{Department of Physics, U.C. Berkeley, Berkeley, CA 94720}
\date{Preprint quant-ph/9501015;
Submitted to Physical Review Letters, 8 June, 1994}
\maketitle
\begin{abstract}
The question in the title may be answered
by considering the outcome of a ``weak measurement'' in the
sense of Aharonov {\it et al.}  Various properties of the resulting
time are discussed, including its close relation to the Larmor times.
It is a universal description of a broad class of measurement interactions,
and its physical implications are unambiguous.
\end{abstract}
\pacs{PACS numbers: 03.65.Bz,73.40.Gk}

The question posed in the title has remained controversial since the early
days of quantum theory
\cite{one-six,%
Buttiker=1982,%
eight-eleven,%
Leavens=1993BK,Landauer=1994}.  One
commonly cited
reason for this is the nonexistence of a quantum-mechanical time operator.
However, it is quite possible to construct an operator $\Theta_B$
which measures
whether a particle is in the barrier region or not.  Such a projection
operator
is Hermitian, and may correspond to a physical observable.
It has eigenvalues 0 and 1, and its
expectation value simply measures the integrated probability density over
the region of interest-- it is this expectation value divided by the incident
flux which is referred to as the ``dwell time''
\cite{Buttiker=1983}.
Thus the central problem is not the absence of an appropriate
Hermitian operator,
but rather the absence of well-defined histories (or trajectories) in
standard
quantum theory.  For
the dwell time measures a property of a wave function with both
transmitted and reflected portions,
and does not display a unique decomposition into portions
corresponding to these individual scattering channels.  Some workers
point out that it can be fruitful to consider
the expectation value for a particular
outgoing channel rather than a particular incident state
\cite{vTandM}.  However, this approach answers the present
title's question no better than does the usual dwell time; instead of
discarding information about late times it discards information about early
times.
Other related approaches follow phase space trajectories
\cite{vTandM},
Bohm trajectories \cite{
Leavens=1993BK},
or Feynman paths \cite{SokFertig=1990}.
No consensus has been reached as to the validity and the
relationship of these various
approaches.  Ideally, transmission and reflection times $\tau_T$ and
$\tau_R$ would, when weighted by the transmission and reflection
probabilities $|t|^2$ and $|r|^2$, yield the dwell time $\tau_d$:
\begin{equation}
\label{weight}
|t|^2\tau_T + |r|^2\tau_R = \tau_d \;\; ;
\end{equation}
this relation has served as one of the main criteria in a broad review of
tunneling times \cite{eight-eleven}, but has also been criticized
 \cite{Landauer=1994}.  In this
paper I present an approach to tunneling times which adheres strictly to
the standard formulation of quantum mechanics, defining the ``dwell''
time by explicitly considering a general von Neumann-style
 measurement interaction.  The
mean value of the measurement outcome may then be calculated
for transmitted and reflected particles
individually, and I show that it automatically
satisfies the above equation. This time is in
general complex, and is closely related to the Larmor times and to the
complex time of Sokolovski, Baskin, and Connor
\cite{SokFertig=1990}; the
consideration of a measuring apparatus, however, makes
interpretation of its real and imaginary parts straightforward,
as will be demonstrated by examining a generalization of the Larmor
time.

Before presenting the new theory, let us recap several
existing approaches.
Relation (\ref{weight}) is violated by many of these approaches,
including perhaps the simplest: the group delay or extrapolated
``phase time.''  This time describes the appearance of a wave packet peak
on the far side of the barrier using the method of stationary phase.
It has been pointed out
that there is no fundamental reason in quantum theory to associate a
``delay'' time of this sort with the duration of the interaction itself
\cite{Buttiker=1982}.
For thick enough
barriers the wave packet may appear to have travelled faster than the
vacuum speed of light {\it c}; this has recently been verified for several
different manifestations of tunneling
\cite{expts}.
Since such effects occur in the limit of very small transmission
probability, the overall ``center-of-mass'' of the wave function ({\it i.e.},
the expectation value of $x$) does not travel faster than light.  The
anomalies occur only when a {\it sub}ensemble is defined both by
preparation of a state incident on the barrier from the left and by
postselection of particles emergent on the right.  In the ``standard''
interpretation of quantum mechanics, this postselection comprises a
collapse of the wave function, which occurs instantaneously
\cite{Heisenberg=1930}, and the apparent superluminal motion of those
particles which are transmitted is due entirely to the instantaneity of this
uncontrollable collapse event, which is useless for signalling purposes.

Another class of approaches to tunneling relies instead on an external
``clock.''  The Larmor time
\cite{B-and-r,Buttiker=1983}, in which a magnetic
field confined to
the barrier region causes the spin of a tunneling particle to precess, is the
prototypical theory of this sort.  As B\"uttiker observed, this
time has two different components, the precession of the spin in the plane
perpendicular to the applied field and the tendency of the spin to align
itself with the field, and so far the choice of one of these components or
their vector sum seems to be merely a matter of opinion.  Several experiments
related to ``clocks''
have been performed \cite{Landauer=1994}.

	Meanwhile, Aharonov {\it et al.} have developed a theory of
``weak measurements'' precisely to deal with questions about
subensembles such as those with which we are concerned
\cite{Aharonov=both}.
Suppose we wish to measure the expectation value of some operator $A$
(for instance, $\Theta_B$)
which acts on the particle under study.  Following von Neumann
\cite{von_Neumann=1955},
let us postulate a measuring apparatus or ``pointer'' whose position and
conjugate momentum we term $Q$ and $P$.  A measurement results from
a time-dependent interaction $H_{int} = g(t) P \cdot A$; since $P$ is the
generator of translations for the pointer, the mean position of the pointer
after the interaction will have shifted by an amount proportional to the
expectation value $\mbox{$\langle A\rangle $}$.  In an ideal measurement, the
relative
shifts corresponding to different eigenvalues of $A$ are large compared
with the initial uncertainty in the pointer's position, and the resulting
lack
of overlap between the final states leads to the effective decoherence
(or irreversible ``collapse'') between different eigenstates of $A$.  In
\cite{Aharonov=both}, this approach is modified
in that the initial position of the pointer has a large uncertainty, so
that the overlap between the pointer states
remains close to unity, and hence that the
measurement does not constitute a collapse.
Seen another way, this means that the pointer
momentum $P$ may be very well-defined, and therefore need not impart
an uncertain ``kick'' to the particle; the measurement is ``weak'' in that it
disturbs the state of the particle as little as possible between the state
preparation and the post-selection.  One can give up measurement
precision in exchange for delicacy.  What Aharonov {\it et al.} calculate is
the average shift of the pointer, according to standard quantum theory, for
those of the particles prepared in state $\mbox{$ | i\rangle $}$ which are
later found to be in state $\mbox{$ | f\rangle $}$.  The derivation can be
found in
\cite{Aharonov=both},
but the same result can be loosely understood from the point
of view of probability theory.
Consider the conditional probability for the particle to be found
in eigenstate $\mbox{$ | A_i\rangle $}$ (with eigenvalue $a_i$)
{\it given} that it is subsequently found to be
in state $\mbox{$ | f\rangle $}$: according to standard Bayesian theory,
$P(A_i | f) = P(A_i \& f) / P(f)$.
The probability $P(f)$ is represented in quantum theory by the
expectation value of the projection operator ${\rm Proj}(f) \equiv
\mbox{$\mbox{$ | f\rangle $}\mbox{$\langle f|$}$}$;
we replace the logical {\it and} with multiplication
since these projectors are binary-valued.  (It
should however be noted that the product of two projectors need not be
Hermitian, so these joint ``probabilities'' are not in general real.
Detailed
discussion of ``extended'' probabilities can be found in
\cite{prob-people}.)  Thus we
rewrite the relation
\begin{equation}
\label{Bayes}
P(A_i | f) = \frac{\mbox{$\langle {\rm Proj}(f){\rm Proj}(A_i)\rangle$}}{
\langle{\rm
Proj}(f)\rangle} \; .
\end{equation}
Summing over $i$, we find the ``weak'' ({\it i.e.}, conditional)
 expectation value
\begin{equation}
\mbox{$\langle A\rangle $}_f = \sum_i a_i \frac{\mbox{$\langle {\rm Proj}(f)
{\rm Proj}(A_i)\rangle$}}
{ \mbox{$\langle {\rm Proj}(f)\rangle$}}
    = \frac{\langle|f\rangle\langle f|A\rangle}{\langle|f\rangle\langle
    f|\rangle} \; .
\end{equation}
Evaluating the expectation values for a particle prepared in $\mbox{$ |
i\rangle $}$, we
find
\begin{equation}
\label{weak}
\mbox{$\langle A\rangle $}_{fi} =
\frac{\langle i|\;|f\rangle\langle f|A \; |i\rangle}{\langle
i|\;|f\rangle\langle f|\;|i\rangle}
=\frac{\langle f|A|i\rangle}{\langle f|i\rangle} \;\; .
\end{equation}
This is the result proved in \cite{Aharonov=both} for the mean shift
in the pointer position.  The symmetry of this expression is
related to the time-reversibility of quantum evolution in the
absence of collapse.
Note that
this expression may in general be complex.  Unlike the case for complex
times found in the past for tunneling, however, the physical significance
of the real and imaginary parts of a weak value is clear.  If the initial
state
of the pointer is $\exp [-Q^2 / 4\sigma^2]$, then after the measurement it
will, for suitably normalized $g(t)$, be in the state
$\exp [-(Q-\mbox{$\langle A\rangle $}_{fi})^2 / 4\sigma^2]$.  The real
part of $\mbox{$\langle A\rangle $}_{fi}$ corresponds to the mean shift in
the pointer
position, while the imaginary part constitutes a shift in the pointer {\it
momentum}.
This latter effect is a reflection of the
back-action of a measurement on the particle.  It
obviously does have physical significance, but since it does not
correspond to a spatial translation of the pointer, should {\it not}
 be thought of
as part of the measurement outcome.  Furthermore, unlike the spatial
translation $\Delta Q = {\rm Re} \mbox{$\langle A\rangle $}_{fi}$, this
effect is sensitive to the
initial state of the pointer: $\Delta P = {\rm Im} \mbox{$\langle A\rangle
$}_{fi} / 2\sigma^2$.
As $\sigma$ becomes large, the measurement becomes very weak, and
the momentum shift of the pointer (like the back-action on the particle)
vanishes, while the spatial shift remains constant.

Consider these weak values for a complete
orthonormal set of final states $\mbox{$ | f_j\rangle $}$.  We take the
weighted
average
\begin{eqnarray}
\overline{\mbox{$\langle A\rangle $}_{fi}}&  \equiv & \sum_j |\mbox{$\langle
f_j|i\rangle $}|^2
\mbox{$\langle A\rangle $}_{f_ji}
 =  \sum_j \mbox{$\langle i|f_j\rangle $}\mbox{$\langle f_j|i\rangle $}
\frac{\mbox{$\langle f_j|$}A\mbox{$ | i\rangle $}}{\mbox{$\langle
f_j|i\rangle $}} \nonumber \\
& = & \sum_j \mbox{$\langle i|f_j\rangle $} \mbox{$\langle f_j|$}A\mbox{$ |
i\rangle $}
 =  \mbox{$\langle i|$}A \mbox{$ | i\rangle $} = \mbox{$\langle A\rangle $}
 \; \; ;
\end{eqnarray}
 thus relations in the form of Eq. (\ref{weight}) are satisfied automatically
if $\tau_T$ and $\tau_R$ are defined in terms of weak values.  This is
because the weak values represent average measurement outcomes for
various subensembles, which must clearly reproduce the average for the
ensemble as a whole when weighted properly.

Let us examine a particle tunneling through a rectangular barrier extending
from $x = -d/2$ to $x = +d/2$.  As explained in the first paragraph, the
operator $A$ we wish to measure is simply the projector onto this region,
$\Theta_B \equiv \Theta(x+d/2)-\Theta(x-d/2)$.  Suppose as above
that we allow a pointer to interact with the particle so that its position
will
reflect the expectation value of this operator.  We prepare the particle
in a stationary state $\mbox{$|i\rangle$}$
defined such that to the right of the barrier, it
contains only right-going components; a superposition of several such
states with a range of energies would describe a wave packet incident
from the left at early times.
\begin{eqnarray}
\label{state}
\psi_i(x<d/2) & = & \exp[+ikx] + r\exp[-ikx] \nonumber \\
\psi_i(-d/2<x<d/2) & = & B\exp[-\kappa x] + C\exp[\kappa x] \nonumber \\
\psi_i(x>d/2) & = & t\exp[+ikx] \; .
\end{eqnarray}
At a later time, we may find the particle in the transmitted state
$\mbox{$|t\rangle$}$ (the time-dependent case is treated explicitly
in \cite{Steinberg=1994PRA}, and in \cite{Falck=1988} for
the Larmor clock.)  In this
event, we observe the pointer position, which constitutes a measurement
of how much time the transmitted particle spent in the barrier region
(compare the definition of dwell time in \cite{Buttiker=1983}).
Over many trials, the average position
is given by Eq. (\ref{weak}), where $\mbox{$ | f\rangle $}$ represents the
state of a
transmitted particle, defined by the stipulation that to the left of
the barrier it contains only right-going components.
This stipulation means that
$\mbox{$ | i\rangle $}$ and $\mbox{$ | t\rangle $}$ are related
by a parity-flip combined with a time-reversal: $\psi_t(x)=\psi_i^*(-x)$.
Similarly, $\psi_r(x)=\psi_i^*(x)=\psi_t(-x)$.  The transition amplitudes
$\mbox{$\langle t|i\rangle $}$ and $\mbox{$\langle r|i\rangle $}$ are simply
$t$ and $r$.
Thus, defining $\tau_f =
\mbox{$\langle \Theta_B\rangle $}_{fi} / J_{\rm in}$, where $J_{\rm in} =
\hbar
k/m$, and using (\ref{state}), we have
\begin{eqnarray}
\label{integrals}
\tau_T & = & \frac{m}{\hbar k}\, \frac{1}{t} \int_{-d/2}^{d/2} dx \;
 \left(Be^{\kappa x} + Ce^{-\kappa x}\right)
 \left(Be^{-\kappa x}+Ce^{\kappa x}\right)
\nonumber \\
\tau_R & = & \frac{m}{\hbar k}\, \frac{1}{r} \int_{-d/2}^{d/2} dx \;
  \left(Be^{-\kappa x} + Ce^{\kappa x}\right)
 \left(Be^{-\kappa x}+Ce^{\kappa x}\right)
\nonumber \\
\tau_d & = & \frac{m}{\hbar k}\, \int_{-d/2}^{d/2} dx \;
 \left|Be^{-\kappa x}+Ce^{\kappa x}\right|^2 \; ,
\end{eqnarray}
Evaluating the integrals, we find
\begin{eqnarray}
\label{times}
\tau_T & = & \frac{m}{\hbar k} \, \frac{1}{t} \left[
(B^2  + C^2) d + (BC+CB) \frac{\sinh \kappa d}{\kappa} \right]
 \nonumber \\
\tau_R & = & \frac{m}{\hbar k} \, \frac{1}{r} \left[
(BC  + CB) d + (B^2+ C^2) \frac{\sinh \kappa d}{\kappa} \right]
\nonumber \\
\tau_d & = & \frac{m}{\hbar k} \,  \left[
(B^*C  + C^*B )d + \left(|B|^2+
|C|^2\right) \frac{\sinh \kappa d} {\kappa} \right]
 \; .
\end{eqnarray}

Since $|B/C| = \exp[\kappa d]$, the dominant contributions to
(\ref{integrals}) are the $B^2$ terms, which decay exponentially
with $x$ for $\tau_R$ and $\tau_d$, reflecting the fact that little
of the wave penetrates far into the barrier.  By contrast, the $B^2$
term is constant across the entire barrier for $\tau_T$.  In a sense,
this suggests that the transmitted particles differ from the others in
that they sample the entire barrier thickness.  However, $\tau_T$ is
predominantly imaginary, and its {\it real part} stems predominantly
from the two extremes of the barrier, becoming independent of
thickness along with the other times.  This feature bears an
interesting resemblance to the intra-barrier group delay time,
which remains nearly constant
 through most of the barrier
(since $B\exp[-\kappa x]$ accumulates no phase and $|B| \gg |C|$)
and grows at a rate of approximately $m/\hbar k$ for the {\it last}
exponential decay length ($1/\kappa$).

It is straightforward (but unenlightening)
 to demonstrate that ${\rm Re}[(B^2+C^2)/t] = {\rm Re}[2BC/r]=2{\rm
 Re}[B^*C]$ and
 ${\rm Re} [2BC/t] = {\rm Re}[(B^2+C^2)/r] =|B|^2+|C|^2$,
which immediately imply the equality of the real parts (that is, the
actual measurable shift of the pointer) for all three times, and thus
trivially
of relation (\ref{weight}).  This holds for all values of $k$ and $d$,
including the classical limit as well as the tunneling regime.
The equality is {\it not} a general property of weak values,
but rather a consequence of (\ref{weight}) and the symmetry of the
barrier.  Since the integrands in (\ref{integrals}) have different
spatial dependences, we see that the conditional dwell times in very
small regions will not be the same for the different scattering
channels, although when integrated over the entire barrier, they
are both equal to the ``unconditional'' dwell time.
The dwell time in regions
outside the barrier depends even more strikingly on which channel
is considered.

The expression for $\tau_T$ is the same one found by
Sokolovski {\it et al.} by performing an appropriately weighted average
over Feynman paths \cite{SokFertig=1990}: its real part
is equal to the dwell time (in the opaque limit, approximately
$mk/\hbar\kappa k_0^2$), as is the in-plane
portion of the Larmor time.  Its imaginary part is equal to minus the
out-of-plane portion of the Larmor time (which reduces in the opaque
limit to the B\"uttiker-Landauer time $md/\hbar\kappa$),
and its magnitude is thus equal to
B\"uttiker's proposed Larmor time.  Due to the emphasis on real measuring
devices, however, the different significance of these portions becomes clear
in the present context:
the real part is the mean shift in the pointer's position, while the
imaginary
part is instead a measure of the back-action on the particle due to the
measurement interaction itself.  In fact, as discussed earlier, this latter
effect
is sensitive to the details of the measurement apparatus (in particular, to
its initial uncertainty in momentum), unlike the position shift.

Although $\tau_T$ calculated in this way is a general description of
{\it any} von Neumann-style measurement, these results can be particularly
well understood in the context of the Larmor time, which is defined
by the interaction between a magnetic field confined to the barrier region
and the particle's spin.  $H_{int} = -\gamma {\bf S \cdot B} =
 -\gamma S_z B_0 \Theta_B$
is nothing more than a particular realization of the von Neumann
interaction.  $S_z$ plays the role of the pointer momentum; as it is the
 difference between the occupation {\it numbers} of the spin-up
and spin-down states, the conjugate pointer position is the difference
between the {\it phases} of the spin-up and spin-down components
 \cite{Dirac=1927}.  This relative phase determines
 the in-plane rotation angle,
which is why the Larmor precession serves as a measure of the dwell in
the barrier region.  The present approach shows that {\it any} interaction
which is designed to measure $\Theta_B$ will yield $\tau_T$, and thus
that the Larmor result is perfectly general, and the natural extension of
standard quantum measurements to experiments which include
postselection.

To see that the tendency of the particle's spin to align with the applied
field
(which B\"uttiker showed dominated over the in-plane rotation, in the
opaque limit) should not be considered an additional term for the traversal
time, we must let the measurement become more ``weak'' by increasing
the initial uncertainty $\sigma$ in the pointer position, as discussed above
after eq. (\ref{weak}).  It might at first seem that it would suffice to
consider particles of spin $\gg 1/2$, in order to approach the
 correspondence principle limit.  B\"uttiker found, however, that the Larmor
time was the same for higher-spin particles.  The various $S_z$ eigenstates
remain split by $\hbar \omega_L \equiv \hbar\gamma B_0$, but if the pointer is
prepared in the maximum-$S_x$ eigenstate (the case typically considered,
in order to follow rotation in the $x-y$ plane), the width of the initial
distribution in $S_z$ is $\sqrt{S/2}$.  For small $\omega_L$, the
transmission varies linearly with $S_z$, so the shift it introduces in
$\mbox{$\langle S_z\rangle $}$ scales as the square of the width (near the
peak, treat
the distribution as quadratic: $[1-(S_z/\Delta S)^2] \cdot [1 + \xi S_z]
\sim 1-[(S_z-\xi\Delta S^2/2)/\Delta S]^2$).  The alignment angle
goes as $S_z/S$ and is thus independent of $S$,
as is the Larmor frequency; the time defined by their ratio is hence
unchanged by going to $S>1/2$.
However, it is possible to
prepare the particle in a ``spin-squeezed state''
\cite{Wineland=1992,Kitagawa=1993}, in which $\Delta S_z$
may be much smaller than $\sqrt{S/2}$, while $\mbox{$\langle \protect\bf
S\rangle $}$
still points along $\bf x$.  The uncertainty-principle trade-off is
that the angle in the $x-y$ plane can no longer be determined as precisely.
The mean rotation angle, however, is determined by the relative phase
difference between successive $S_z$-states, and this is not affected by
the weighting of the different components.  As the Larmor measurement
is made weaker by using a highly squeezed initial spin-state, the in-plane
rotation (the real part of $\tau_T$) remains constant, while the out-of-plane
rotation (the imaginary part of $\tau_T$) vanishes.  Thus while the former
timescale is indeed a property of the tunneling process, regardless of how
it is observed, the latter is merely a measure of the back-action provoked
by the measurement, sensitive to the initial state of the measurement
apparatus.

It follows from the above argument that ${\rm Re}\, \tau_T$ is in essence
a derivative of the transmission phase with respect to potential energy
(since
the Larmor Hamiltonian is merely a spin-dependent potential), while
${\rm Im}\, \tau_T$ is a derivative of the transmission amplitude.  It is
well
known
that the phase shifts for reflection and for transmission given a spatially
symmetric barrier differ only by a constant, leading to equal group delay
times for the two scattering channels \cite{Falck=1988,reciprocity}.  This
is also why ${\rm Re}\, \tau_R = {\rm Re}\, \tau_T$, and hence (taking
(\ref{weight})
into account) why both times are equal to $\tau_d$.  (In fact, $\tau_g
\approx \tau_d$ except at low energies.)  Similarly, it is the
fact that $|r|^2+|t|^2=1$ which determines the behavior of the derivatives
of the amplitudes, ensuring that $|t|^2{\rm Im}\, \tau_T = - |r|^2{\rm Im}\,
\tau_R$.

It might be asked whether ${\rm Re}\, \tau_T$ alone represents the entire
duration
of the tunneling interaction, or whether ``self-interference delays'' prior
to
the barrier constitute an additional contribution.  When the
three times discussed in this paper are generalized to treat regions other
than the barrier, all three
do contain interference terms, but the resulting spatial
oscillations continue indefinitely, leading to no overall contribution when
integrated over all the space to either side of the barrier.  It is
interesting to
note that while the dwell time displays such oscillations only to the left of
the barrier, the other times oscillate everywhere outside the barrier.  In
particular, the ``reflection dwell time'' at points
$x$ to the right of the barrier
$d\tau_R (x > d/2) \propto \sin (2kx + \arg\, t) dx$.
It may seem odd that the reflected particles display a non-zero dwell time on
the far side of the barrier, but the physical interpretation is clear.  Any
interaction which can detect the presence of a particle (including, but not
restricted to, an applied magnetic field) has some probability of reflecting
the particle.  In the Larmor case, it is easy to show
 that the spin-up and spin-%
down particles will be reflected with equal probability but opposite phase
by a small magnetic field at some point beyond the barrier.  Some of these
particles are transmitted back to the left side of the barrier, and the phase
difference means that on average, the spin angle of particles emerging
on the left now
depends on the strength of the applied field on the far side of the barrier.
One may be tempted to argue that the effect comes about due to particles
 which would not have been reflected in the absence of a measurement,
and hence
is not truly indicative of the behavior of reflected particles in general;
in particular, \cite{Falck=1988} and \cite{Leavens=1994PRA} conclude that
such times are unphysical.
If, however, we follow the Copenhagen interpretation, we have
no choice but to define our question in terms of conceivable measurements.
This leads automatically to $\tau_T$ and $\tau_R$ as found in this paper,
including their odd properties beyond the barrier.  While the back-action
 on the particle may vary with measurement technique, the delays
calculated in this fashion do not.
As discussed in \cite{Steinberg=1994PRA},
for example, the momentum transfer due to the Coulomb interaction between a
proton and an electron may become repulsive.  Whether or not one describes
this as a negative interaction time, it is a counter-intuitive and
physically testable consequence of these conditional expectation
values.

This work was supported by ONR grant N00014-90-J1259.  Discussions
with R. Chiao, A. Elby, P. Kwiat, and M. Mitchell were helpful.

\end{document}